\documentclass[runningheads]{llncs}

\usepackage[T1]{fontenc}
\def\doi#1{\href{https://doi.org/\detokenize{#1}}{\url{https://doi.org/\detokenize{#1}}}}
\usepackage{graphicx}
\usepackage{amsmath}
\usepackage{newtxmath}
\usepackage{amsfonts}
\usepackage{multirow}
\usepackage{tabularx}
\usepackage{ulem}
\usepackage{bm}
\usepackage[dvipsnames]{xcolor}
\usepackage{threeparttable}
\usepackage{amsmath}
\usepackage[pagebackref=true,breaklinks=true,colorlinks,bookmarks=false]{hyperref}

\usepackage{pdfpages}

\usepackage{booktabs}

\usepackage{color}

\usepackage{listings}
\begin{document}
\title{Regression Metric Loss: Learning a Semantic Representation Space for Medical Images}
%
%
\author{
Hanqing Chao
\and Jiajin Zhang
\and Pingkun Yan 
\thanks{Corresponding author.} 
}


%
\authorrunning{H. Chao et al.}
%
\institute{Department of Biomedical Engineering and 
Center for Biotechnology and Interdisciplinary Studies,
Rensselaer Polytechnic Institute, Troy, NY, USA 12180\\
\email{\{chaoh, zhangj41, yanp2\}@rpi.edu}}

\titlerunning{Regression Metric Loss}

\maketitle              
\begin{abstract}
Regression plays an essential role in many medical imaging applications for estimating various clinical risk or measurement scores. While training strategies and loss functions have been studied for the deep neural networks in medical image classification tasks, options for regression tasks are very limited. One of the key challenges is that the high-dimensional feature representation learned by existing popular loss functions like Mean Squared Error or L1 loss is hard to interpret. In this paper, we propose a novel Regression Metric Loss (RM-Loss), which endows the representation space with the semantic meaning of the label space by finding a representation manifold that is isometric to the label space. Experiments on two regression tasks, \textit{i.e.} coronary artery calcium score estimation and bone age assessment, show that RM-Loss is superior to the existing popular regression losses on both performance and interpretability. Code is available at \url{https://github.com/DIAL-RPI/Regression-Metric-Loss}.

\end{abstract}
\section{Introduction}
Various clinical risk or measurement scores, such as coronary artery calcium (CAC) score~\cite{agatston1990quantification}, bone age~\cite{gilsanz2005hand}, etc., have been widely used in clinical practice for quantifying the progression of diseases and pathological conditions~\cite{oprita2014scores}. 
Tasks for regressing such scores, \textit{i.e.} estimating the continuous variables from medical images, are an important group of applications in medical image analysis~\cite{litjens2017survey}. 
Similar to other medical imaging applications like image segmentation and computer-aided diagnosis, where deep learning has reshaped the landscape of research by continuously pushing the performance to new record levels, deep neural networks (DNNs) have also been applied to regressing clinical scores~\cite{gilsanz2005hand,liu2019multi,dai2021adaptive}.

However, compared with those relatively direct applications of DNNs for predicting discrete class labels, medical image regression is a more challenging task for DNNs due to the nature of predicting continuous quantity. 
Compared with the abundant options available for the classification tasks, choices for regression tasks are very limited.
A majority of existing deep learning models for medical image regression are trained with Mean Squared Error (MSE) or L1 loss~\cite{dai2021adaptive}. Although these models might have satisfactory performance, the high-dimensional feature representations learned by these deep models are hard to interpret. 
Recent studies on classification tasks have shown that imposing appropriate constraints can force deep learning models to learn more semantically meaningful representations~\cite{ye2020augmentation,choi2020amc,zhang2021explainable}. 
A few works on medical image regression has tried to transplant popular distance metric learning losses such as N-pair loss~\cite{sohn2016improved} and triplet loss~\cite{hermans2017defense} by introducing an adaptable margin~\cite{zheng2021semi,dai2021adaptive}. 
However, such adaptations lack theoretical foundations and cannot fully explore the inherent structure of the continuous label space. 
We argue that to learn a meaningful representation space for regression, a loss should be able to explore the structure of the continuous label space and reveal such structure in the representation space but not limited to changing the margin in loss functions.

In this paper, we propose a novel loss function suitable for regression problems, named Regression Metric Loss (RM-Loss).
It guides the deep learning model to learn a low-dimensional manifold that has the same semantic meaning as the label, in the high-dimensional feature representation space. 
Specifically, the RM-Loss explicitly constrains the learned manifold and the label space to be isometric, \textit{i.e.}, the geodesic distance between the representations of any two data points on the manifold is proportional to the Euclidean distance between their labels. 
Thus, each point on this manifold can be interpreted by the corresponding point in the label space. Assigning such a semantic meaning to the feature representation manifold is of the essence for label regression, which provides direct interpretability for the regression methods. With such a semantic representation, at the time of inference, the regression result of an input test sample can be computed based on its distance to the nearest neighbors on the training manifold, instead of using some uninterpretable fully connected layers for mapping its representations to an output value.
In addition, since our proposed RM-Loss constrains the distance between data pairs, it automatically augments the number of training samples and helps the model explore relationships between samples. This in turn helps boost the model performance, especially on small datasets.

The key contributions of this paper are as follows. 1) We propose a novel loss for medical image regression tasks, the Regression Metric Loss (RM-Loss). It constrains a DNN model to learn a semantically meaningful manifold
that is isometric to the label space.
2) At the inference time, the regression result is determined by the distance-weighted nearest neighbors. By providing the exact neighbor samples and their weights as an interpretation, such an approach would be more informative for clinical applications than only outputting a final score. 
3) Experimental results on two clinical regression tasks over two large-scale public datasets, CAC score estimation from computed tomography images and bone age regression from X-ray images, show that the proposed RM-Loss achieved superior performance and interpretability.

\section{Method}

This section presents the details of our proposed RM-Loss. We aim to help a DNN learn a low-dimensional manifold in the high-dimensional feature representation space, which has the same semantic meaning as the label space.

\subsection{Learning a Semantically Interpretable Representation Space}

Let $\{(\bm{x}_i, \bm{y}_i)\}_{i=1}^N$ denotes a regression dataset of size $N$, where $\bm{x}_i$ is an input sample, for instance an image, and $\bm{y}_i \in \mathbb{R}^{d_y}$ is a $d_y$-dimensional label vector. We assume that the label space $Y$ is a Euclidean space. Let $F: \bm{x}_i \mapsto \bm{f}_i$ represents a deep learning model that maps an input sample to a $d_f$-dimensional representation $\bm{f}_i$. The RM-Loss constrains the model such that all the representations $\{\bm{f}_i\}_{i=1}^N$ are on a Riemannian manifold $M$ and the geodesic distance $G(\cdot,\cdot)$ between a pair of samples' representations $\bm{f}_i$ and $\bm{f}_j$ is proportional to the Euclidean distance between their corresponding labels $\bm{y}_i$ and $\bm{y}_j$. 
Thus, the key term for RM-Loss to minimize is
\begin{equation}
    \label{eq:g_iso}
    D^o_{ij} = \left|s \times G(\bm{f}_i, \bm{f}_j) - ||\bm{y}_i-\bm{y}_j||_2\right|,
\end{equation}
where $s$ is a learnable parameter that will be optimized together with all other parameters in the deep learning model $F$.
Eq.~\ref{eq:g_iso} constraints the learned manifold $M$ to be isometric to the label space $Y$. Furthermore, since $M$ is a Riemannian manifold and $Y$ is a Euclidean space, $M$ is diffeomorphic to $Y$. It indicates that the learned manifold $M$ will have the same topological structure as the label space $Y$. Therefore, we can interpret the high-dimensional feature representations on the manifold $M$ by using the corresponding labels.

\subsection{Local Isometry: from Geodesic to Euclidean Distance}
\label{sec:m_local}

The key challenge to minimize Eq.~\ref{eq:g_iso} is to calculate the geodesic distance $G(\cdot, \cdot)$ on the manifold $M$. Below we state a useful lemma of the isometry between a Riemannian manifold and a Euclidean space\footnote{The derivation of this lemma is provided in the Supplementary Material.}. 
\begin{lemma}
\label{lem:lg}
Let $M$ be a connected complete Riemannian manifold, and $E$ be a Euclidean space. If a differentiable mapping $h: M \to E$ is a local isometry, we have $h$ as a global isometry.
\end{lemma}
By Lemma~\ref{lem:lg}, assuming the learned Riemannian manifold $M$ is connected and complete, then as long as $M$ is locally isometric to the label space $Y$, it will be diffeomorphic to $Y$. Furthermore, in the local neighborhood of $\bm{f}_i$ on $M$, $\mathcal{N}(\bm{f}_i)$, we approximate the geodesic distance $G(\bm{f}_i, \bm{f}_j), \bm{f}_j\in\mathcal{N}(\bm{f}_i)$ by the Euclidean distance, 
$||\bm{f}_i-\bm{f}_j||_2$. 
In practice, instead of searching for neighbors of each point $\bm{f}_i$, we calculate $D_{ij}$ for all sample pairs in a training batch and weight them by a Gaussian function according to their labels:
\begin{equation}
  \begin{aligned}
    \label{eq:w}
    D_{ij} = \left| \right. s  \times ||\bm{f}_i -\bm{f}_j||_2 - & ||\bm{y}_i-\bm{y}_j||_2 \left. \right|,
    ~ w_{ij} = \exp \left(-\frac{||\bm{y}_i-\bm{y}_j||_2^2}{2\sigma^2} \right) + \alpha, \\
    & l' = \frac{\sum_{i=1}^N\sum_{j=1}^Nw_{ij}D_{ij}}{\sum_{i=1}^N\sum_{j=1}^Nw_{ij}},
  \end{aligned}
\end{equation}
where $l'$ shows the objective function of current RM-Loss, and $\sigma$ and $\alpha$ are two hyper-parameters. The $\sigma$ controls the size of the neighborhood. 
The $\alpha$ controls the curvature of the manifold to avoid self tangling, \textit{i.e.}, data points away from each other on the manifold will not be too close in the Euclidean space $\mathbb{R}^{d_f}$.
We use the distance between labels instead of representations in $w_{ij}$, because at the beginning of training, the representations are randomly mapped to the $\mathbb{R}^{d_f}$ space. Using the label distance in $w_{ij}$ guides the model to map the representations to a right manifold. When the right manifold is found, the $w_{ij}$ can well depict the neighborhood of $\bm{f}_i$.

\subsection{Hard Sample Pairs Mining}

Similar to most of the distance metric loss, in the training process, the deep model $F$ can quickly learn to correctly map the easy samples, rendering a large fraction of $D_{ij}$ close to zero. These small terms will dramatically dilute the averaged loss value and slow down the training. Thus, we introduce a technique for mining the hard sample pairs to help the training focus on informative sample pairs. Specifically, only the sample pairs, whose weighted losses are greater than the batch average value, will be selected through a masking process as:
\begin{align}
    m_{ij} = 
    \begin{cases} 
    1, & \mbox{if~}w_{ij}D_{ij} > \bar{l}^{(k)} \\
    0, & \mbox{if~}w_{ij}D_{ij} <= \bar{l}^{(k)}
    \end{cases}~,~~
    \bar{l}^{(k)} = 0.9 \times \bar{l}^{(k-1)} +  0.1 \times \mathbb{E}^{(k)}(w_{ij}D_{ij}),
\end{align}
where $k$ indicates the $k$-th iteration in the training, and $\bar{l}$ is the exponential moving average of the mean loss values of all the sample pairs in one training batch. Thereby, the full objective function of the RM-Loss is formulated as:
\begin{equation}
\label{eq:full}
    \mathcal{L} = \frac{\sum_{p=1}^N\sum_{q=1}^Nm_{ij}w_{ij}D_{ij}}{\sum_{p=1}^N\sum_{q=1}^Nm_{ij}w_{ij}}.
\end{equation}

\subsection{Making a Grounded Prediction}

Because the learned manifold $M$ is diffeomorphic to the label space $Y$, each point $f_i$ on $M$ and its neighborhood $\mathcal{N}^M(f_i) \subseteq M$ correspond to a unique point $y_i$ on $Y$ and a neighborhood $\mathcal{N}^Y(y_i) \subseteq Y$, respectively. This property assures that samples with similar representations (in $\mathcal{N}^M(f_i)$) have similar labels (in $\mathcal{N}^Y(y_i)$). Therefore, we can use nearest neighbors (NN) to estimate the label for a test sample. We assume a neighborhood $\mathcal{N}^M(f_i)$ on $M$ can be approximately regarded as a Euclidean space. Since the distance on $M$ has specific semantic meaning defined by the label space, we apply distance-weighted NN within a fixed radius $r$
\footnote{The algorithm proposed for efficient NN radius optimization for model selection is described in Supplementary Material.}
. For a test sample $t$, its label $\hat{y}_t$ can be computed by:
\begin{equation}
\label{eq:inference}
    \hat{y}_t = \frac{\sum_{f_i\in\mathcal{N}_r(f_t)}a_i y_i} {\sum_{f_i\in\mathcal{N}_r(f_t)}a_i}~,~~
    a_i = \exp\left (-\frac{||\bm{f}_i-\bm{f}_t||_2^2}{2(r/3)^2} \right )
\end{equation}
where $\mathcal{N}_r(f_t)$ is the neighborhood of the test sample's representation $f_t$ in the Euclidean representation space $\mathbb{R}^{d_f}$ with a radius $r$. $f_i$ and $y_i$ denote the representations and labels of the training samples in $\mathcal{N}_r(f_t)$.

\section{Experiments}

We evaluate the proposed RM-Loss from two aspects, \textit{i.e.}, the regression performance and the quality of the learned space.
The regression performance is measured by the \textbf{Mean Absolute Error} (MAE) and \textbf{R-squared} (R$^2$). The quality of the learned space is quantified by:
1) The \textbf{Averaged Rank-5 Difference} (D5) calculated by the label differences between each test sample and its 5-nearest neighbors from training samples. 2) The \textbf{Residual Variance} (RV)~\cite{tenenbaum2000global} defined as: $1-\rho(G_M, D_Y)$, where $\rho(\cdot)$ is the Pearson correlation coefficient, $G_M$ is the geodesic distance of all test sample pairs on the learned representation manifold $M$, and $D_Y$ is the label Euclidean distance of the same set of test sample pairs. $G_M$ is calculated following the original paper~\cite{tenenbaum2000global} by $k$-nearest neighborhoods. 
The $k$ leads to the best (smallest) RV is selected. This measurement directly evaluates whether the manifold $M$ and the label space $Y$ are isometric. 

\subsection{Experimental Settings}
\subsubsection{Datasets}
Two public datasets, RSNA Bone Age Assesment Dataset (BAA)~\cite{halabi2019rsna} and NLST CAC score estimation dataset (CAC), are used for evaluation.
\textbf{BAA Dataset}: Since the labels of original validation and test set of BAA dataset are not publicly available. We randomly split its original $12,371$ training images into training, validation, and test sets in a proportion of $7:1:2$. The input images are all resized into $520\times400$. In the training phase, the labels ($\in [4,228]$) are z-score normalized.
\textbf{CAC Dataset}: This dataset is originated from the National Lung Screening Trial (NLST) dataset~\cite{nlst_2011}. It contains $43,241$ low dose chest CT images from $10,395$ subjects. These subjects are randomly split into training ($7299$, $70\%$, $30,347$ images), validation ($1066$, $10\%$, $4,365$ images), and test sets ($2,030$, $20\%$, $8,529$ images). The heart region of each image is first segmented by a pre-trained U-Net~\cite{ronneberger2015u}. The CAC in the heart region is further segmented by another U-Net. Finally, a CAC score (Agatston score~\cite{agatston1990quantification}) is calculated based on the size and intensity of the CAC. According to the clinical grading standard\footnote{Agatston score: minimal: 1-10, mild: 11-100, moderate: 101-400, severe: >400}, the $\log_2$CAC scores ($\in [0, 13]$) are used as labels in the following experiments. In the training phase, the labels are scaled into $[-3.5,3.5]$ ($(\mbox{label}-6.5)/2$). The segmented heart region is cropped out and resampled into $128\times128\times128$ with a resolution of 1.6mm$\times$1.6mm$\times$1.07mm.

\noindent\textbf{Baselines}
We compared our proposed RM-Loss with widely used MSE loss, L1 loss, ordinal regression loss (OrdReg) (225 heads for the BAA dataset; 26 heads for the CAC dataset), and the recently proposed adaptive triplet loss (ATrip)~\cite{zheng2021semi} (margin$=1$). For the adaptive triplet loss, since it is applied on the representation space, in its original paper~\cite{zheng2021semi}, it has to be applied together with L1 or MSE loss for final prediction. In our experiment, for fair comparison, we present the results of both ATrip+L1 and ATrip+NN, where NN is the same nearest neighbors algorithm used in our method. 
All the losses are compared on the exactly same baseline models. For the BAA dataset, the winning method of the RSNA Pediatric Bone Age Machine Learning Challenge~\cite{halabi2019rsna} was used, which is a Inception Net V3~\cite{szegedy2016rethinking}. For the CAC dataset, a 3D-ResNet is applied~\cite{hara2017learning}. Detailed structures of these two models are shown in the Supplementary Material.

\noindent\textbf{Implementation Details}
For all experiments, Adam with a learning rate of $1e-4$ is applied as the optimizer. \textbf{BAA Dataset}: Unless otherwise stated, the $\sigma$ and $\alpha$ in $w_{ij}$ (Eq.~\ref{eq:w}) are set to be $0.5$ and $0.1$, respectively. The $r$ in Eq.~\ref{eq:inference} is calculated as $0.162$ by the algorithm described in Supplementary S-Fig 1. All the models are trained for $15,000$, $10,000$ iterations for the full and $10\%$ size dataset\footnote{Details of using partial dataset are described in Sec.~\ref{sec:data_eff}}, respectively, with a batch size of $64$. \textbf{CAC Dataset}: Unless otherwise stated, the $\sigma$ and $\alpha$ are set to be $1.5$ and $0.1$ respectively. The $r$ in Eq.~\ref{eq:inference} is calculated as $0.291$. All the models are trained for $30,000$, $10,000$ iterations for the full and $10\%$ size dataset, respectively, with a batch size of $36$.
Our source code will be released on GitHub.

\subsection{Main Results}

Tab.~\ref{tab:main_res} shows the regression performance and the quality of the learned representation space on the BAA and CAC datasets. The proposed RM-Loss achieved both best regression performance and highest space quality. These results support our claim that, 
the proposed RM-Loss can guide the model learn a representation manifold that have similar topological structure as the label space, and the predictions made based on this interpretable manifold are accurate. 
As a distance metric loss ATrip+NN achieved better performance than L1 on the CAC dataset, but not on the BAA dataset.
A possible reason is that ATrip loss only controls the relative distance among anchor, positive, and negative triplets. For a classification task, this is enough to force the representations to cluster by groups. However, for a regression task, where the label space is continuous, this constrain cannot give the model a clear direction to optimize. Therefore, sometimes it can find a good representation but sometimes it cannot. The performances of ATrip+NN are better than ATrip+L1 on both datasets demonstrate the importance of using distance-weighted NN in the inference time. When applied together with L1, a distance metric loss cannot perform at its best.

\begin{table}[t]
	\centering
	\caption{Regression performance and quality of the learned space on the full-size BAA and CAC datasets. The best results are in bold.
	}
	\label{tab:main_res}
\begin{threeparttable}
	\begin{tabular}{|l||c|c|c|c||c|c|c|c|}
    \hline
    \multirow{3}{*}{Method} & \multicolumn{4}{c||}{Regression Performance} & \multicolumn{4}{c|}{Space Quality} \\
    \cline{2-9}
    & \multicolumn{2}{c|}{BAA} & \multicolumn{2}{c||}{CAC} & \multicolumn{2}{c|}{BAA} & \multicolumn{2}{c|}{CAC} \\
    \cline{2-9}
    & MAE$\downarrow$ & R$^2\uparrow$ & MAE$\downarrow$ & R$^2\uparrow$ & D5$\downarrow$  & RV$\downarrow$  & D5$\downarrow$  & RV$\downarrow$  \\
    \hline
    MSE      & ~7.021$^*$ & ~0.947 & ~0.762$^*$ & ~0.930 & ~8.872$^*$ & ~0.0690$^*$ & ~0.869$^*$ & ~0.1225$^*$ \\
    \hline
    L1       & ~6.580$^*$ & ~0.952 & ~0.668$^*$ & ~0.927 & ~8.799$^*$ & ~0.0600~ & ~0.837$^*$ & ~0.1249$^*$ \\
    \hline
    OrdReg   & ~7.061$^*$ & ~0.944 & ~0.684$^*$ & ~0.937 & ~9.368$^*$ & ~0.1450$^*$ & ~0.844$^*$ & ~0.1966$^*$ \\
    \hline
    ATrip+NN & ~6.799$^*$ & ~0.951 & ~0.647$^*$ & ~0.939 & ~9.022$^*$ & ~0.0617~ & ~0.834$^*$ & ~0.0874$^*$ \\
    \hline
    ATrip+L1 & ~6.854$^*$ & ~0.949 & ~0.660$^*$ & ~0.930 & ~9.055$^*$ & ~0.0630$^*$ & ~0.855$^*$ & ~0.0813~ \\
    \hline
    RM-Loss (ours)~~~ & ~\textbf{6.438} & ~\textbf{0.954} & ~\textbf{0.617} & ~\textbf{0.944} & ~\textbf{8.614} & ~\textbf{0.0588} & ~\textbf{0.769} & ~\textbf{0.0806} \\
    \hline
	\end{tabular}
\begin{tablenotes}\footnotesize
\item[$*$] $p$<0.05 in the one-tailed paired \textit{t}-test. The significant test was not applied on $R^2$.
\end{tablenotes}
\end{threeparttable}
\end{table}

\begin{figure}[t]
	\centering
	\includegraphics[width=\textwidth]{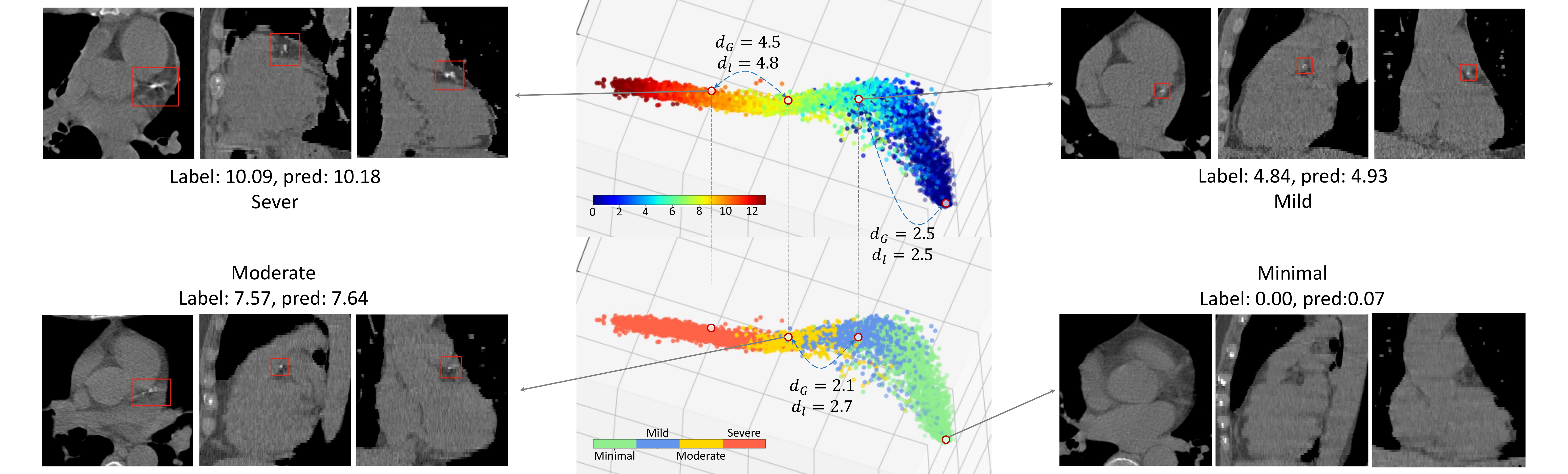}
\caption{Visualization of the learned representation space on the CAC dataset. Linear PCA was used for reducing the dimensionality to 3D for visualization. Two color maps are applied on the representations: a continuous color map indicating the original label (top), and a discrete color map shows the CAC score grading results (bottom). $d_G$ and $d_l$ represents the geodesic distance between representations and the Euclidean distance between labels, respectively.}	
\label{fig:vis}
\end{figure}

To better understand the relationship between the learned representation space and the label space, we visualize the learned space and the corresponding sample cases in Fig.~\ref{fig:vis}. The geodesic distance of the four sample cases is highly correlated with their label distance. The visualization of the space learned by other loss functions are included in the Supplementary Material.

\subsection{Ablation Studies}

\begin{table}[t]
	\centering
	\caption{Ablation study of $\sigma$, $\alpha$, and $m$ on BAA dataset. The best results are in bold.}
	\label{tab:abl}
	\begin{tabular}{|l||c|c|c|c|c||c|c|c||c|}
    \hline
    $\sigma$ & 0.25 & 0.5 & 1.0 & 1.5 & $+\infty$ &  \multicolumn{4}{c|}{$0.5$}  \\
    \hline
    $\alpha$ & \multicolumn{5}{c||}{$0.1$} & 0.0 & 0.2 & 0.3 & 0.1 \\
    \hline
    $m$ &  \multicolumn{8}{c||}{w/} & w/o \\
    \hline
    MAE$\downarrow$& 6.555 & \textbf{6.438} & 6.642 & 6.759 & 6.823 & 6.496 & 6.591 & 6.741 & 6.520 \\
    \hline
    R$^2\uparrow$  & 0.953 & \textbf{0.954} & 0.952 & 0.951 & 0.950 & \textbf{0.954} & 0.953 & 0.951 & 0.953 \\
    \hline
    D5$\downarrow$ & 8.726 & \textbf{8.614} & 8.905 & 8.930 & 9.096 & 8.707 & 8.875 & 8.786 & 8.717 \\
    \hline
    RV$\downarrow$ & 0.0614& \textbf{0.0588}& 0.0658& 0.0659& 0.0699& 0.0650& 0.0614& 0.0677& 0.0641 \\
    \hline
	\end{tabular}
\end{table}

In Tab.~\ref{tab:abl}, we studied the influence of the two hyper-parameters ($\sigma$ and $\alpha$) and the hard sample pair mining on the BAA dataset. Similar results are observed on the CAC dataset, which are included in the Supplementary Material.

$\bm{\sigma}$:~In Eq.~\ref{eq:w}, $w_{ij}$ is calculated on the labels $y_i$ and $y_j$. So the hyperparameter actually have the same semantic meaning as the label. Based on the role of $w_{ij}$, $\sigma$ controls the size of region on the manifold.
Since two Euclidean spaces can be mapped to each other with a linear mapping, $\sigma$ indicates the degree that the representation space can be linearly related to the label space. A too large $\sigma$ would decrease the non-linearity of the learned manifold such that its representation ability will be decreased, while a too small  $\sigma$ would cause overfitting. In Tab.~\ref{tab:abl}, $\sigma=+\infty$ is equivalent to remove $w_{ij}$ from Eq.~\ref{eq:full}, so that the model will tend to learn a linear space. 
It is worth noting that, our empirically found optimal $\sigma$s on both two datasets are pretty close to the labels' diagnostic boundaries. In clinical practice, children with bone age more than 2 years (24 months) advanced or delayed are recommended to get a further diagnosis. By scaling with the standard deviation $41.2$ we used on BAA dataset, a $\sigma=0.5$ indicates a bone age difference of $20.6$ months which is very close to 24 months. On the CAC dataset, since the label is log scaled, a $\sigma=1.5$ indicates a $8$ times difference, while according to the standard of Agatston score, a $10$ times difference will increase or decrease one grade.
$\bm{\alpha}$:~The $\alpha$ also adjusts the linearity of the learned manifold, but in a global way. As shown in Tab.~\ref{tab:abl}, an increase of $\alpha$ would cause a dramatic drop of the performance. However, a relatively small $\alpha$ can constrain the non-linearity of the manifold to avoid overfitting. Thus, we used a consistent $\alpha=0.1$ on both two datasets.
$\bm{m}$:~The last column in Tab.~\ref{tab:abl} shows that removing hard sample pair mining would cause a performance decrease.

\begin{table}[t]
	\centering
	\caption{The regression performance and the learned space quality on the $10\%$-size BAA and CAC dataset with the best results in bold.
	}
	\label{tab:10p}
\begin{threeparttable}
	\begin{tabular}{|l||c|c|c|c||c|c|c|c|}
    \hline
    \multirow{3}{*}{Method} & \multicolumn{4}{c||}{Regression Performance} & \multicolumn{4}{c|}{Space Quality} \\
    \cline{2-9}
    & \multicolumn{2}{c|}{$10\%$ BAA} & \multicolumn{2}{c||}{$10\%$ CAC} & \multicolumn{2}{c|}{$10\%$ BAA} & \multicolumn{2}{c|}{$10\%$ CAC} \\
    \cline{2-9}
    & MAE$\downarrow$ & R$^2\uparrow$ & MAE$\downarrow$ & R$^2\uparrow$ & D5$\downarrow$  & RV$\downarrow$  & D5$\downarrow$  & RV$\downarrow$  \\
    \hline
    MSE      & ~8.721$^*$ & ~0.917 & ~0.946$^*$ & ~0.895 & ~11.204$^*$ & ~0.1054$^*$ & ~1.102$^*$ & ~0.1706$^*$  \\
    \hline
    L1       & ~9.173$^*$ & ~0.906 & ~0.875$^*$ & ~0.894 & ~11.682$^*$ & ~0.1133$^*$ & ~1.028$^*$ & ~0.1529$^*$   \\
    \hline
    OrdReg   & ~9.226$^*$ & ~0.908 & ~0.849$^*$ & ~0.906 & ~11.821$^*$ & ~0.2485$^*$ & ~1.010$^*$ & ~0.2189$^*$ \\
    \hline
    ATrip+NN & ~8.733$^*$ & ~0.911 & ~0.861$^*$ & ~0.907 & ~10.990~ & ~0.1018$^*$ & ~1.056$^*$ & ~0.1547$^*$ \\
    \hline
    ATrip+L1 & ~9.017$^*$ & ~0.914 & ~0.875$^*$ & ~0.908 & ~11.208$^*$ & ~0.1016$^*$ & ~1.012$^*$ & ~0.1142~  \\
    \hline
    RM-Loss (ours)~~~ & ~\textbf{8.071} & ~\textbf{0.926} & ~\textbf{0.797} & ~\textbf{0.912} & ~\textbf{10.974} & ~\textbf{0.0878} & ~\textbf{0.971} & ~\textbf{0.1114}\\
    \hline
	\end{tabular}
\begin{tablenotes}\footnotesize
\item[$*$] $p$<0.05 in the one-tailed paired \textit{t}-test. The significant test was not applied on $R^2$.
\end{tablenotes}
\end{threeparttable}
\end{table}

\subsection{Data Efficiency}
\label{sec:data_eff}

In this section, we compare the performance of different losses when the training data is limited by using only $10\%$ of the training and validation sets to train the models. The testing sets remain in full size. Tab.~\ref{tab:10p} shows that our proposed RM-Loss achieved much better regression performance than the other losses, due to its efficiency in exploring relationships between data pairs.

\section{Conclusion}

In this work, we proposed a novel distance metric loss originated for medical image regression tasks named as Regression Metric Loss (RM-Loss). By constraining data samples to have geodesic distances in the representation space corresponding to their Euclidean distance in the label space, the RM-Loss guides deep models to learn a semantically interpretable manifold. Experiments on two different regression tasks shows that the models trained with RM-Loss achieve both superior regression performance and high quality representation space. \\

\noindent \textbf{Acknowledgements.} This work was partly supported by National Heart, Lung, and Blood Institute (NHLBI) of the National Institutes of Health (NIH) under award R56HL145172.

\bibliographystyle{splncs04}
\bibliography{refs}



\section*{Supplementary material}

# Regression Metric Loss: Learning a Semantic Representation Space for Medical Images (Supplementary Material)

## 1 Proof of Lemma 1

Given $M$ be a connected complete Riemannian manifold (feature space), $E$ a connected Euclidean manifold (label space) and $h : M \to E$ a differentiable mapping that is locally an isometry. $h$ is a global isometry if and only if it satisfies the injectivity and surjectivity propoerties.

**Injectivity** Let $p \neq q \in M$ and let $\gamma : I \to M$ be a geodesic satisfying $\gamma(0) = p$ and $\gamma(1) = q$. Cover $\gamma(I)$ by open sets $U_\alpha$ where $h|_{U_\alpha} : U_\alpha \to h(U_\alpha)$ is an isometry. By compactness, there is a finite covering of $U_i$'s or equivalently, a partition of $I$ where $0 = t_0 < t_1 < ... < t_n = 1$. Thus, $\frac{D}{dt}(\frac{dh\circ\gamma}{dt}|_{[t_i,t_{i+1}]}) = dh\frac{D}{dt}(\frac{\gamma}{dt}|_{[t_i,t_{i+1}]}) = dh(0) = 0$. Therefore, $\frac{D}{dt}(\frac{dh\circ\gamma}{dt})$ on $I$ implying that $h \circ \gamma$ is a geodesic in $E$ joining $h(p)$ to $h(q)$. If $h(p) = h(q)$, then $h \circ \gamma$ would be a closed geodesic, contradicting the uniqueness assumption.

**Surjectivity** Uniqueness of geodesics joining any two points of manifold $E$ implies that $E$ is complete, then $exp_q : T_q N \to E$ is surjective for any $q \in E$. Let $p \in M$ be fixed. There is an open neighborhood $U \in M$ containing $p$ s.t. $h|_U : U \to h(U)$ is an isometry. We have $h(exp_p(v)) = exp_{h(p)}(dh_p(v))$ for all $v \in T_p M$. Let $q \in N$ be arbitrary. There is a $w \in T_{h(p)} N$ s.t. $exp_{h(p)}(w) = q$. Since, $dh_p$ is an isomorphism of vetor spaces, there is a $u \in T_p M$ s.t. $dh_p(u) = w$. Hence, $h(exp_p(v)) = exp_{h(p)}(dh_p(v)) = q$, $h$ satisfies the surjectivity.

Thus, $h$ is a global isometry.

## 2 Efficient NN Radius Optimization for Model Selection

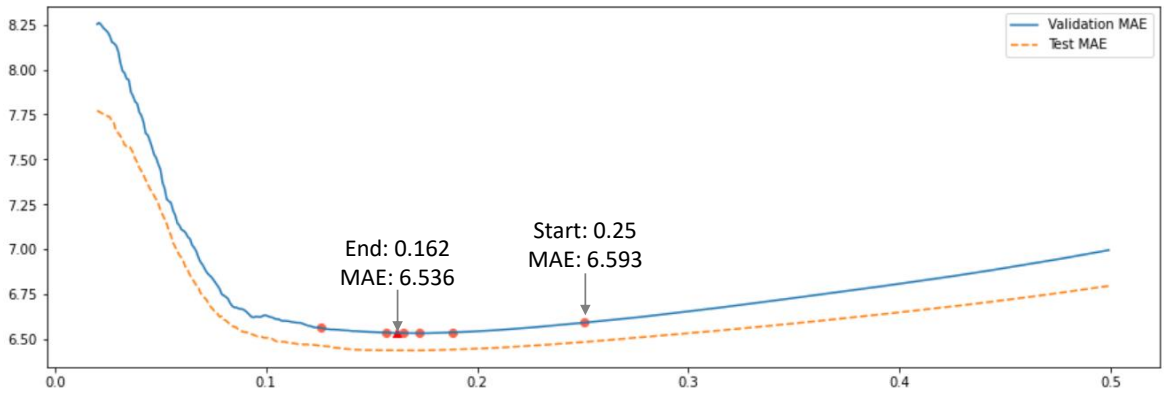

S-Figure 1: The radius $r$ significantly influences the performance of the NN algorithm. Although one can conduct grid search in a feasible range, it would be time consuming especially for the checkpoint selection of a deep learning model. We observed that the relationship between the performance and radius $r$ always have a trend similar as a convex function. So we use the performance on $r$ and $r + \epsilon$ to approximate the gradient of the performance on $r$, and apply a binary search to find a local minimum with gradient near to zero. With a proper $\epsilon$ (=0.01 in our experiments), this local minimum will be very close to the global minimum.

## 3 Model Architectures

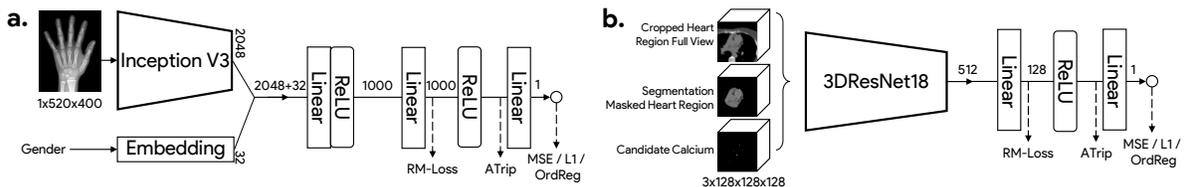

S-Figure 2: Architectures of Networks used in experiments. **a.** Inception V3 used for the BAA dataset. **b.** 3D-ResNet18 used for the CAC dataset.

1

# 4 Visualization of the Learnt Representation Spaces

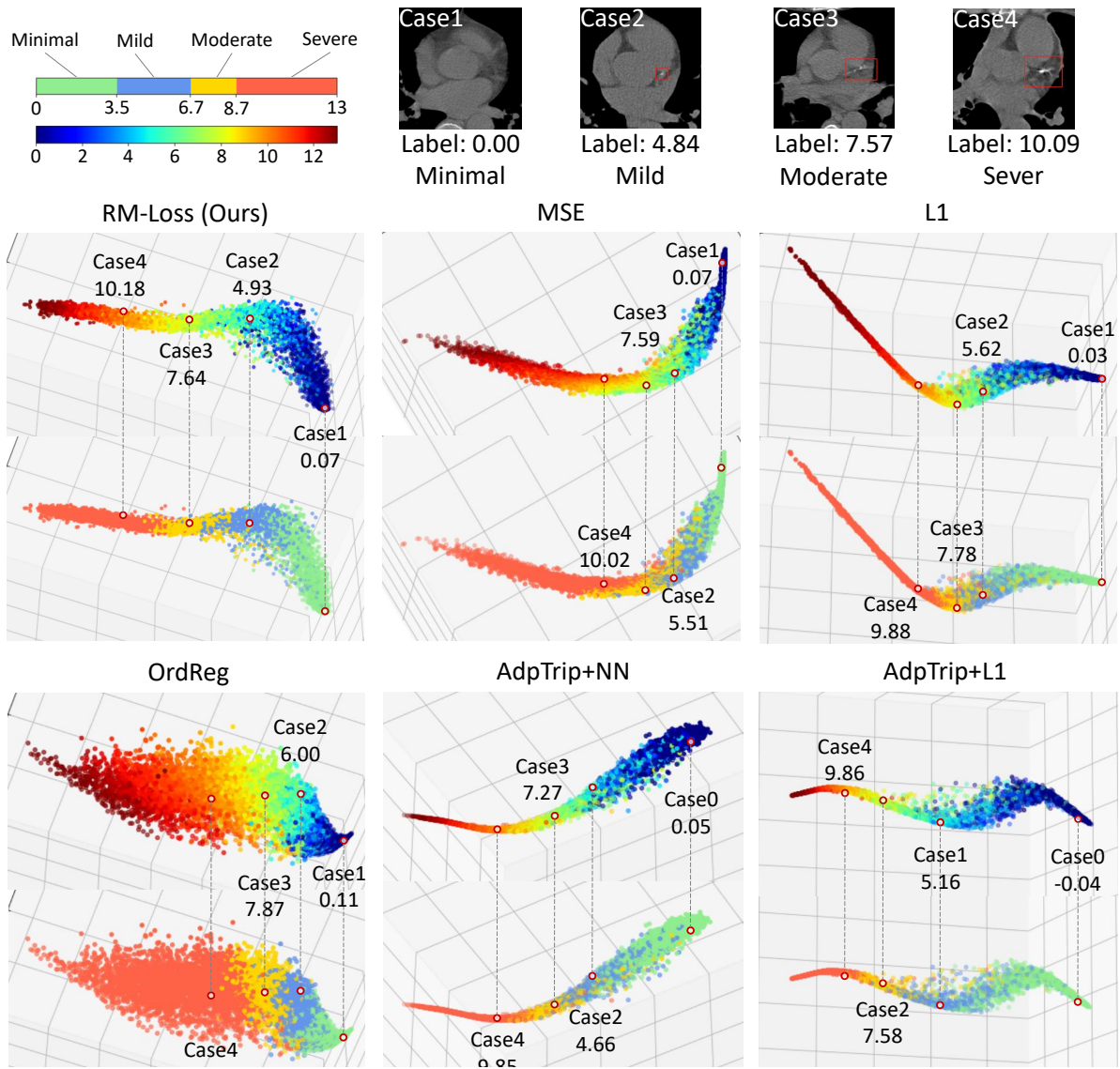

S-Figure 3: Visualization of all baselines and our RM-Loss on the CAC dataset. The numbers in the plots are models' predictions on the 4 sample cases shown on the top row.

# 5 Ablation Study on the CAC dataset

S-Table 1: Ablation study of $\sigma$, $\alpha$, and $m$ on the full and $10\%$ size CAC, and $10\%$ size BAA dataset.

| $m$ | $\alpha$ | $\sigma$ | CAC MAE↓ | R²↑ | D5↓ | RV↓ | 10% CAC MAE↓ | R²↑ | D5↓ | RV↓ | $\sigma$ | 10% BAA MAE↓ | R²↑ | D5↓ | RV↓ |
|---|---|---|---|---|---|---|---|---|---|---|---|---|---|---|---|
| w/ | 0.1 | 0.5 | 0.635 | 0.943 | 0.800 | 0.0771 | 0.779 | 0.920 | 0.948 | 0.1029 | 0.25 | 8.091 | 0.925 | 11.024 | 0.0879 |
|  |  | 1.0 | 0.630 | 0.941 | 0.796 | 0.0833 | 0.812 | 0.913 | 0.978 | 0.1134 | 0.5 | 8.071 | 0.926 | 10.974 | 0.0878 |
|  |  | 1.5 | 0.617 | 0.944 | 0.769 | 0.0806 | 0.797 | 0.912 | 0.971 | 0.1114 | 1.0 | 8.489 | 0.920 | 11.199 | 0.0955 |
|  |  | 2.0 | 0.646 | 0.939 | 0.813 | 0.0788 | 0.840 | 0.905 | 1.025 | 0.1617 | 1.5 | 7.958 | 0.926 | 10.797 | 0.8635 |
|  |  | $+\infty$ | 0.614 | 0.942 | 0.762 | 0.0897 | 0.840 | 0.911 | 1.001 | 0.1153 | $+\infty$ | 8.379 | 0.922 | 11.314 | 0.0940 |
|  | 0.0 | 1.5 | 0.626 | 0.943 | 0.772 | 0.1042 | 0.801 | 0.910 | 0.996 | 0.1195 | 0.5 | 9.743 | 0.895 | 12.504 | 0.1216 |
|  | 0.2 |  | 0.636 | 0.943 | 0.788 | 0.1054 | 0.853 | 0.910 | 0.982 | 0.1136 |  | 8.544 | 0.915 | 11.855 | 0.1015 |
|  | 0.3 |  | 0.637 | 0.941 | 0.797 | 0.0829 | 0.843 | 0.910 | 0.997 | 0.1354 |  | 8.216 | 0.922 | 11.144 | 0.0931 |
| w/o | 0.1 |  | 0.591 | 0.946 | 0.755 | 0.0722 | 0.838 | 0.915 | 0.987 | 0.1208 |  | 8.699 | 0.915 | 11.651 | 0.1015 |


\end{document}